\documentclass[aps,prb,twocolumn,showpacs,floats]{revtex4-1}
\usepackage[dvips]{graphicx}
\usepackage{amsmath}
\usepackage{amssymb}
\usepackage{empheq}
\usepackage{bm}
\newcommand{\eps}{\epsilon}

\newcommand{\ra}{\rangle}
\newcommand{\la}{\langle}
\newcommand{\ud}{\mathrm{d}}

\begin{document}

\title{\bf Magnetic texture-induced thermal Hall effects}

\author{Kevin A. van Hoogdalem,$^1$ Yaroslav Tserkovnyak,$^2$ and Daniel Loss$^1$}

\affiliation{$^{1}$ Department of Physics, University of Basel, Klingelbergstrasse 82, CH-4056 Basel, Switzerland}
\affiliation{$^{2}$Department of Physics and Astronomy, University of California, Los Angeles, California 90095, USA}

\date{\today}

\begin{abstract}
Magnetic excitations in ferromagnetic systems with a noncollinear ground state magnetization experience a fictitious magnetic field due to the equilibrium magnetic texture. Here, we investigate how such fictitious fields lead to thermal Hall effects in two-dimensional insulating magnets in which the magnetic texture is caused by spin-orbit interaction. Besides the well-known geometric texture contribution to the fictitious magnetic field in such systems, there exists also an equally important contribution due to the original spin-orbit term in the free energy. We consider the different possible ground states in the phase diagram of a two-dimensional ferromagnet with spin-orbit interaction: the spiral state and the skyrmion lattice, and find that thermal Hall effects can occur in certain domain walls as well as the skyrmion lattice.
\end{abstract}

\maketitle
\section{Introduction}
Traditionally, the role of information carrier in spin- and electronic devices is taken by respectively the spin or the charge of the conduction electrons in the system. In recent years, however, there has been an increasing awareness that spin excitations in insulating magnets (either magnons or spinons) may be better suited for this task. The reason behind this is that these excitations are not subject to Joule heating. Therefore, the energy associated with the transport of a single unit of information carried by a magnon (or spinon) current can be much lower in such insulating systems.~\cite{Trauz08} The quasiparticle bosonic nature of the magnons, furthermore, allows in principle to essentially eliminate losses due to scattering or contact impedances at low temperatures.~\cite{Mei03}

Creation of a magnon current has been shown to be possible by means of the spin Seebeck effect,~\cite{Uch10} the spin Hall effect,~\cite{Kaj10} and with high spatial accuracy by means of laser-controlled local temperature gradients.~\cite{Wei12} The resulting spin current can be measured utilizing the inverse spin Hall effect.~\cite{Kaj10,San11} It has been shown that the magnon current can propagate over distances of several centimeters in yttrium iron garnet (YIG).~\cite{Kaj10} It has recently been shown that it is theoretically possible to implement the analogs of different electronic components using insulating magnets.~\cite{Mei03,Hoo11} 

Hall effects for magnon currents are of interest both from a fundamental point of view as well as from the point of view of applications. Even though the physical magnetic field does not directly couple to the orbital motion of neutral magnons, certain kinds of spin-orbit interactions can lead to Hall phenomena similar to those of a charged particle in a magnetic field. Mechanisms that have been shown to give rise to nonzero Hall conductances in certain insulating magnets include coupling of spin chirality to a magnetic field~\cite{Kat10} and Dzyaloshinskii-Moriya (DM) interaction.~\cite{Ono10} Of interest for applications is the fact that Hall effects in insulating magnets allow one to control the magnon spin current.

\begin{figure}
\centering
\includegraphics[width=0.95\columnwidth]{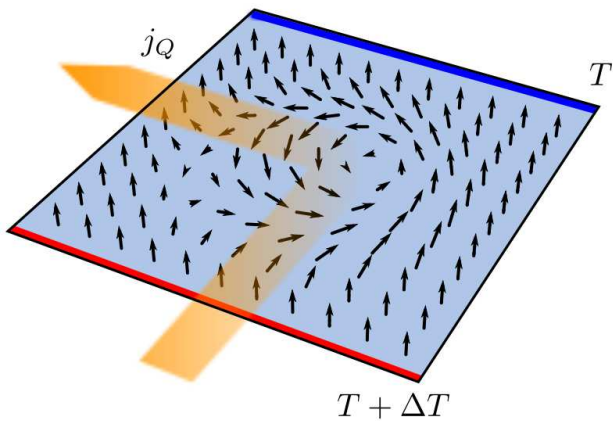}
\caption[]{Pictorial representation of the thermal spin Hall effect. A temperature difference $\Delta T$ applied to a sample leads to a finite heat current. Since the heat current is carried by the magnons in the system, the fictitious magnetic field that magnons experience due to a non-trivial magnetic ground state will lead to a finite thermal Hall conductivity.}
\label{fig:Fig1}
\end{figure}

Recently, Katsura {\it et al.} predicted~\cite{Kat10} a nonzero thermal Hall conductivity (see Fig. \ref{fig:Fig1}) for the Heisenberg model on the Kagom\'e lattice. The finite conductivity originates from the fact that the coupling of spin chirality to an applied magnetic field leads to a fictitious magnetic flux for the magnons in the specific case of the Kagom\'e lattice. Later, Onose {\it et al.} measured~\cite{Ono10} the thermal Hall effect in the pyrochlore ferromagnet Lu$_2$V$_2$O$_7$. In this experiment, the combination of DM interaction and the pyrochlore structure leads to the finite thermal Hall conductivity.

In those previous studies, the thermal Hall effect was considered using a quantum-mechanical lattice model as starting point. The symmetry of the underlying lattice played a crucial role. We take a different approach and consider insulating ferromagnets with a noncollinear ground-state magnetic texture, which we model using a phenomenological description. It is well known that the effect of the presence of a noncollinear ground state on the elementary excitations in a ferromagnet can be captured by introducing a fictitious electromagnetic potential in the equation of motion for the magnons.~\cite{She04,Dug05} Spin-orbit interactions generally also contribute non-Abelian gauge fields into the magnetic Hamiltonian.~\cite{Li11} Furthermore, nonlinearized gauge fields for Dzyaloshinskii-Moriya interaction were derived in Ref. \onlinecite{Han10} using the CP$^1$ representation. There are, correspondingly, two contributions to the fictitious electromagnetic potential. The first one depends only on the equilibrium magnetic texture; the second depends on the form of the free energy that gives rise to the noncollinear ground state in the first place, {\it i.e.}, the contribution to the free energy due to spin-orbit interaction. Since both contributions are determined by the spin-orbit interaction, they will generally be of similar magnitude.

It has been shown that the fictitious electromagnetic potential couples the motion of magnetic texture and that of heat currents.~\cite{Kov09} This coupling reveals itself through local cooling by magnetic texture dynamics~\cite{Kov09} and thermally induced motion of magnetic textures such as domain walls.~\cite{Hat,Ber}

This work is organized as follows. In Sec. \ref{Sec:II} we introduce our system and derive the fictitious electromagnetic vector potential that acts on the magnons, which turns out to include an often-overlooked contribution. In Sec. \ref{Sec:III} we derive the relevant ground state properties of the different ground states in the phase diagram of an insulating ferromagnet with nonzero DM interaction. In Sec. \ref{Sec:IV} we calculate the band structure of one of the ground states, the triangular skyrmion lattice, and calculate its thermal Hall conductivity.

\section{Magnons in the presence of magnetic texture}
\label{Sec:II}
We consider a two-dimensional nonitinerant ferromagnet in the $x$-$y$ plane with spatially varying and time-dependent spin density $s{\bf m}({\bf r},t)$. The spin density is related to the magnetization ${\bf M}({\bf r},t)$ as  $s{\bf m}({\bf r},t) = {\bf M}({\bf r},t)/\gamma$, where $\gamma$ is the gyromagnetic ratio ($\gamma < 0$ for electrons). The magnitude $s$ of the spin density is assumed to be constant, and ${\bf m}({\bf r},t)$ is a unit vector. The system is described by the Lagrangian~\cite{She04,Dug05}
\begin{equation}
 \mathcal{L} = \int \ud^2 {\bf r}\left[ {\bf D}({\bf m}) \cdot \dot{{\bf m}}-F({\bf m}, \partial_j {\bf m})\right].
\label{eq:L}
\end{equation}
Here ${\bf D} = s\hbar\left( {\bf n}_\textrm{D} \times {\bf m}\right)/(1+{\bf m} \cdot {\bf n}_\textrm{D})$ is the vector potential corresponding to the Wess-Zumino action with an arbitrary ${\bf n}_\textrm{D}$ pointing along the Dirac string. $F({\bf m}, \partial_j {\bf m})$ is the magnetic free energy density of the system,  which we assume to be of the form (here $j=x,y$; double indices are summed over)
\begin{equation}
 F({\bf m}, \partial_j {\bf m}) =\frac{Js}{2} \left(\partial_j {\bf m}\right)^2 - M_s {\bf m} \cdot {\bf H} + s F_{\Gamma}({\bf m},\partial_j {\bf m}).
\label{eq:F}
\end{equation}
Here $J$ is the strength of the exchange interaction, $M_s =\gamma s$ is the saturation magnetization, ${\bf H}$ the external magnetic field (which we will always assume to be in the $z$ direction), and $F_{\Gamma}({\bf m},\partial_j {\bf m})$ describes terms due to broken symmetries. For isotropic ferromagnets in the exchange approximation, the leading-order terms in the free energy are quadratic in the texture [first term in Eq. (\ref{eq:L})]. Breaking inversion symmetry by spin-orbit interactions, while still retaining isotropy in the $x$-$y$ plane, allows us to construct terms that are first order in texture. These terms are given by
\begin{equation}
F_{\Gamma}({\bf m},\partial_j {\bf m}) = \Gamma_{\textrm{R}} m_z\nabla\cdot {\bf m} + \Gamma_{\textrm{DM}} {\bf m} \cdot \left(\nabla \times {\bf m}\right).
\label{eq:FGamma}
\end{equation}
We defined $\nabla = \partial_x{\bf x} + \partial_y{\bf y}$, where ${\bf x},{\bf y}$ are unit vectors. The first term is due to structural inversion symmetry breaking and hence is anisotropic in the $z$ direction. Such terms occur in systems with finite Rashba spin-orbit interaction~\cite{Bych84} or on the surface of a topological insulator.~\cite{Tse11} The second term describes DM interaction,~\cite{DM58} which originates from the breaking of bulk inversion symmetry and is therefore isotropic. We note that the two terms in Eq. (\ref{eq:FGamma}) are equivalent (up to an irrelevant boundary term) under a simple rotation around the $z$ axis in spin space. Since such a rotation does not have any additional effect on the equation of motion for the magnetization, Eq. (\ref{eq:LLeq}), we can always absorb the term proportional to $\Gamma_{\textrm{R}}$ in the term proportional to $\Gamma_{\textrm{DM}}$. We will therefore put $\Gamma_{\textrm{R}}$ to zero in the remainder of this work.  For simplicity, we have ignored a term $-\kappa m_z^2$ that would describe easy axis anisotropy, and a term $-M_s {\bf m} \cdot {\bf H_m}/2$, where ${\bf H_m}$ describes the magnetic stray field, in Eq. (\ref{eq:F}).

Substitution of Eq. (\ref{eq:L}) in the Euler-Lagrange equation leads to the Landau-Lifshitz equation
\begin{equation}
s \hbar \dot{{\bf m}} - {\bf m} \times \delta_{{\bf m}} \mathcal{F}({\bf m}, \partial_j {\bf m}) = 0,
\label{eq:LLeq}
\end{equation}
where $ \mathcal{F}({\bf m}, \partial_j {\bf m})$ is the total magnetic free energy of the system. We split the magnetization ${\bf m}$ in a static equilibrium magnetization ${\bf m}_0$ and small fast oscillations $\delta{\bf m}$ (spin waves) around the equilibrium magnetization. To lowest order in $\delta {\bf m}$ the two are orthogonal. In a textured magnet ${\bf m}_0 = {\bf m}_0 ({\bf r})$, which makes finding the elementary excitations a nontrivial task. To circumvent this issue we introduce a coordinate transformation ${\bf m}'({\bf r}) = \hat{R}({\bf r}){\bf m}({\bf r})$, where $\hat{R}({\bf r})$ is such that the new equilibrium magnetization ${\bf m}_0'$ is constant and parallel to the $z$ axis. In this coordinate frame the spin waves are in the $x$-$y$ plane.

The $3\times 3$ matrix $\hat{R}$ describes a local rotation over an angle $\pi$ around the axis defined by the unit vector ${\bf n} = \left[{\bf z} + {\bf m}_0\right]/\left[2\cos \left(\theta/2\right)\right]$. Here, $\theta$ is the polar angle of ${\bf m}_0$, and ${\bf z}$ is a unit vector. Using Rodrigues' rotation formula, we find $\hat{R} = 2{\bf n} {\bf n}^T - \hat{1}$. The effect of the transformation to the new coordinate system is that we have to use the covariant form of the differential operators, $\partial_\mu \to (\partial_\mu +\hat{A}_\mu)$, with $\hat{A}_\mu = \hat{R}^{-1}(\partial_\mu\hat{R})$, in the Landau-Lifshitz equation. The subscript $\mu$ describes both time ($\mu = 0$) and space ($\mu = 1,2$) coordinates.

In the new coordinate system, the Landau-Lifshitz equation for the free energy Eq. (\ref{eq:F}) becomes
\begin{equation}
 i \hbar\dot{m}_+ =\left[J \left(\nabla/i + {\bf A}\right)^2 + \varphi\right] m_+.
\label{eq:vonNeumann}
\end{equation}
 Here, $m_{\pm} = (\delta m_x' \pm i\delta m_y')/\sqrt{2}$ describes circular spin waves in the rotated frame of reference. Furthermore, $\varphi = {\bf m}_0\cdot {\bf H}/s + \hbar[\hat{R}^{-1}(\partial_t\hat{R})]|_{12}$, and the components of the vector potential ${\bf A}$ are given by $A_j= \hat{\mathcal{A}}_j|_{12}$. The skew-symmetric matrices $\hat{\mathcal{A}}_j$ are here defined as $\hat{\mathcal{A}}_j = \hat{R}(\partial_j -\zeta\hat{I}_j)\hat{R}$. In the latter equation we defined $\zeta = \Gamma_{\textrm{DM}}/J$, and the matrices
\begin{equation}	
 \hat{I}_x = \left(\begin{array}{ccc} 0 & 0 & 0 \\ 0 & 0 & -1 \\ 0 & 1 & 0 \end{array}\right) \textrm{ and } \hat{I}_y = \left(\begin{array}{ccc} 0 & 0 & 1 \\ 0 & 0 & 0 \\ -1 & 0 & 0 \end{array}\right).
\end{equation}
We see that the components $A_j$ of the fictitious magnetic vector potential consist of two contributions. The first comes from the exchange interaction in the presence of magnetic texture; the second (texture-independent) part originates from the DM interaction term in the free energy. While it may be tempting to neglect the latter contribution, we will show here that it has important consequences. Indeed, typically both contributions will be of the same order of magnitude. This is because the magnetic texture itself is caused by the DM interaction, and will therefore manifest itself on lengthscales $J/\Gamma_{\textrm{DM}}$. 

We can quantize Eq. (\ref{eq:vonNeumann}) by introducing the bosonic creation operator $b^{\dagger} \propto m_-$. This quantization works since $m_+'$ and $m_-'$ satisfy approximate bosonic commutation relations in the limit of small deviations from equilibrium. After quantization, Eq. (\ref{eq:vonNeumann}) can be interpreted as the von Neumann equation belonging to the Hamiltonian
\begin{equation}
 \mathcal{H} = \int \ud^2 {\bf r} b^\dagger \left[J \left(\nabla/i + {\bf A}\right)^2 + \varphi\right]b.
\label{eq:Hbos}
\end{equation}
Therefore, the elementary excitations of the system behave as noninteracting bosonic quasiparticles. The effect of the smoothly varying equilibrium magnetization is captured by the inclusion of a fictitious magnetic vector potential ${\bf A}$ and electric potential $\varphi$. 

In the derivation of Eq. (\ref{eq:vonNeumann}) we have assumed that the length of a typical wave packet is much smaller than the spatial extension over which the magnetic texture varies. We will refer to this as the adiabatic approximation.~\cite{Kov12} Using this assumption, we have neglected terms in Eq. (\ref{eq:vonNeumann}) that are higher order in texture. Such terms, which become important at lower wave vectors, lead to two distinct effects.~\cite{Kov12} Firstly, a term $-J[(\hat{\mathcal{A}}_{x}|_{13})^2+(\hat{\mathcal{A}}_{x}|_{23})^2+(\hat{\mathcal{A}}_{y}|_{13})^2+(\hat{\mathcal{A}}_{y}|_{23})^2]/2$, which is quadratic in magnetic texture, has to be added to the fictitious electric potential $\varphi$ in Eq (\ref{eq:vonNeumann}) at low wave vectors. Secondly, at low wave vectors one needs to add to the right-hand side of Eq. (\ref{eq:vonNeumann}) a term $J[(\hat{\mathcal{A}}_{x}|_{13}+i\hat{\mathcal{A}}_{x}|_{23})^2+(\hat{\mathcal{A}}_{y}|_{13}+i\hat{\mathcal{A}}_{y}|_{23})^2]m_-$, which introduces a finite ellipticity of the magnons.
 
\section{Textured ground states}
\label{Sec:III}
In this section we present  the different possible ground states for systems with free energy given by Eq. (\ref{eq:F}) (with $\Gamma_{\textrm{R}} = 0$), as a function of the external magnetic field ${\bf H} = H {\bf z}$. We also present the fictitious magnetic vector potentials that find their origin in these textured ground states. It has been shown~\cite{Bog94,Han10} that as the magnetic field $H$ increases from zero, the ground state of a two-dimensional ferromagnet with spin-orbit interaction changes from a spiral state for $H < H_{c1}$, to a skyrmion lattice state for magnetic fields $H_{c1} < H < H_{c2}$, and finally to the ferromagnetic ground state for $H > H_{c2}$. Both critical fields $H_{c1}$ and $H_{c2}$ are typically of the order $\Gamma_{\textrm{DM}}^2/J$ (see Refs. \onlinecite{Han10,Yu10}). This last observation, in combination with the adiabatic assumption and the fact that the equilibrium magnetization is time independent, allows us to neglect the fictitious electric potential $\varphi$ in Eq. (\ref{eq:Hbos}). Since the ferromagnetic ground state has no magnetic texture, it is of no interest for our purposes. In this section we will therefore derive the properties of the spiral and skyrmion lattice ground state.
\subsection{Spiral state}
Following Ref. \onlinecite{Bog94} we will derive the properties of the spiral ground state ${\bf m}_0({\bf r})$ of a two-dimensional ferromagnet with DM interaction. We write ${\bf m}_0({\bf r})$ in the following form
\begin{equation}
{\bf m}_0({\bf r}) = \cos \xi\sin\theta {\bf x} + \sin \xi \sin \theta {\bf y} + \cos\theta{\bf z}.
\label{eq:gen_mag}
\end{equation}
For the spiral state, $\theta = \theta(y)$ and $\xi$ is a constant. With these constraints, the free energy becomes a functional that depends only on $\theta(y)$ and $\partial_y \theta(y)$. Minimizing this functional with respect to $\theta(y)$ gives the following differential equation,
\begin{equation}
\partial_y^2\theta +\alpha \sin \theta = 0,
\label{eq:pend}
\end{equation}
where we defined $\alpha = -\gamma H/J$. Equation (\ref{eq:pend}) is the equation of motion for the mathematical pendulum. The general solution is given in implicit form by the expression
\begin{equation}
\int_0^{\theta(y)}\frac{\ud \theta}{2}\frac{1}{\sqrt{1-m^2\sin^2\theta/2}} = \frac{1}{2}\beta y, 
\end{equation}
where $m = 4 \alpha/(2\alpha+C)$ and $\beta = \sqrt{2\alpha+C}$.  Alternatively, we can write $\theta(y) = 2\phi(\beta y/2,m)$, where $\phi(u,m)$ is the amplitude of the Jacobi elliptic function. The constant $C$ is the first constant of integration. To determine it, we use the fact that $\theta(y)$ is a periodic function with period $y_0$. By integrating the inverse of the first integral $\partial_y \theta$ of Eq. (\ref{eq:pend}) over one period we can determine $y_0$ as
\begin{equation}
y_0 = \int_0^{2\pi} \ud \theta \frac{1}{\sqrt{2\alpha \cos \theta + C}}.
\end{equation}
To fix $C$, we minimize the average free energy $(1/y_0)\int_0^{y_0}F(\theta,\partial_y\theta)$, which leads to the following implicit expression for $C$
\begin{equation}
\int_0^{2\pi} \ud \theta \sqrt{2\alpha \cos\theta + C} = 2\pi \zeta.
\label{eq:impC}
\end{equation}
The minimization of the average free energy also fixes $\cos \xi = 1$. From this we see that the ground state is a spiral state whose structure locally resembles a Bloch domain wall, as is expected for the DM interaction.~\cite{Bog94} We also note that in the case of zero magnetic field ($\alpha = 0$) the spiral state is described by a simple sinusoid with period $y_0 = 2\pi/\zeta$, whereas for finite magnetic field the mirror symmetry with respect to the $x-y$ plane is  broken. Equation (\ref{eq:impC}) also puts a constraint on the maximum value of $H$ for which the spiral state is stable.

\begin{figure}
\centering
\includegraphics[width=0.99\columnwidth]{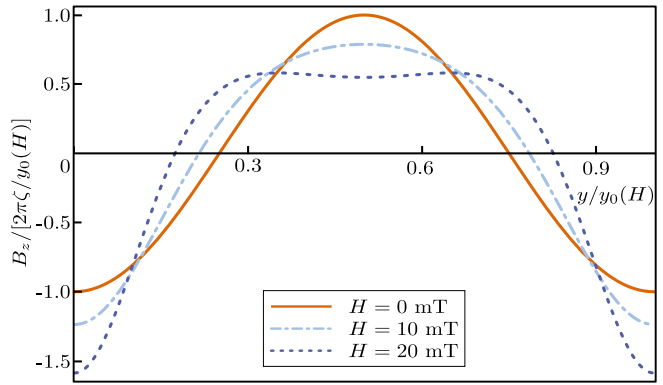}
\caption[]{Fictitious magnetic field due to the spiral ground state. Parameters are $\zeta = 70$ $\mu$m$^{-1}$, $J/(k_B\eps^2) = 63$ K, and the interatomic spacing is taken to be $\eps = 4.5$ \AA. (See Ref. \onlinecite{Yu10}) To make the connection to electromagnetism, we note that a fictitious field $B_z = 2\pi \zeta/y_0(0)\approx 5\cdot 10^{15}$ m$^{-2}$ acting on a spin wave gives rise to the same magnetic length as a $\frac{\hbar}{e}\zeta^2 \approx 3$ T magnetic field acting on a free electron.}
\label{fig:Fig2}
\end{figure}

Some general observations can be made with regard to the fictitious magnetic vector potential due to the spiral ground state. For the ground state Eq. (\ref{eq:gen_mag}) with $\theta = \theta(y)$ and $\xi$ constant, the fictitious vector potential is ${\bf A}(y) = \zeta \sin\theta(y){\bf x}$. This potential is caused solely by the DM contribution to ${\bf A}$; the geometric texture contribution is zero everywhere. The $z$ component of the fictitious magnetic field that the magnons experience is given by $B_z(y) = \left.\nabla \times {\bf A} \right|_z = -\zeta \partial_y \theta(y) \cos \theta(y)$. It is easily seen that the total fictitious magnetic flux over one period of the spiral $\la B_z \ra = \int_0^{y_0} \ud y B_z(y) = 0$. The fictitious magnetic field $B_z(y)$ has been plotted in Fig. \ref{fig:Fig2} for different magnitudes of the applied magnetic field ${\bf H}=H{\bf z}$. Transport in the presence of a magnetic field that is spatially varying in one direction and has zero average has been studied extensively (see Ref. \onlinecite{Nog10} for a recent review). It is well known that these systems do not display a finite Hall conductivity. However, such magnetic fields have been predicted to influence the longitudinal conductance, due to the presence of localized snake orbits at energies that are low compared to the cyclotron frequency associated with the amplitude of the magnetic field.~\cite{Ibr95,Mat00} From our analysis it is also seen that one-dimensional textures can give rise to a nonzero average fictitious magnetic flux for certain domain walls, since these consist of only half a period of the spiral. Hence, such domain walls will display the thermal Hall effect.

Lastly, we note that a proper statistical mechanical description of the spiral phase in three dimensions (or less) requires the inclusion of leading-order non-linearities in the free energy.~\cite{Rad11} The role of those non-linearities in the thermal Hall physics is yet to be understood.
\subsection{Skyrmion lattice}
For magnetic fields $H_{c1} < H < H_{c2}$ the ground state of the two-dimensional ferromagnet with DM interaction is a skyrmion lattice.~\cite{Yu10} This triangular lattice has basis vectors ${\bf a}_1 = a{\bf x}$ and ${\bf a}_2 = (a/2){\bf x} + (a\sqrt{3}/2){\bf y}$, and contains skyrmions with radius $R$. The size of a single unit cell is $(\sqrt{3}/2)a^2$, where $a = 2R$. The magnetization ${\bf m}_0({\bf r})$ of a single skyrmion of radius $R$ centered at the origin is parametrized in polar coordinates $(\rho,\phi)$ by Eq. (\ref{eq:gen_mag}) with $\theta = \theta(\rho)$ and boundary conditions
\begin{equation}
 \theta(0) = \pi \textrm{ and } \theta(R) = 0.
\end{equation}
Furthermore, $\xi = N\phi-\pi/2$, where $N$ is the charge of the skyrmion. We will assume $N=1$ throughout. The magnetization profile can in principle be determined numerically by minimizing the free energy with the aforementioned boundary conditions. However, for simplicity we will assume a linear dependence $\theta(\rho) = \pi(1-\rho/R)$ for our analysis of the texture-induced thermal Hall effect.

In polar coordinates the fictitious magnetic vector potential ${\bf A}({\bf r})$ due to a single skyrmion centered at the origin is given by (here $0\leq \rho \leq R$ and ${\bm \phi}$ is a unit vector)
\begin{equation}
{\bf A} ({\bf r})=  \left[\frac{\cos \theta(\rho)-1}{\rho} - \zeta \cos \theta (\rho)\right]{\bm \phi}.
\label{eq:Askyr}
\end{equation}
The $z$ component of the fictitious magnetic field for this vector potential is given by $B_z(\rho) = \rho^{-1}\partial_\rho (\rho A_\phi)$. It follows that the total flux through a unit cell is $ \la B_z \ra = 2\pi\int_0^R \ud \rho \rho B_z = 4\pi$. This means that each unit cell contains two magnetic flux quanta. The nonzero average flux is caused by the texture contribution to ${\bf A}({\bf r})$; the DM-interaction contribution averages to zero. From the fact that the average magnetic flux is nonzero, it follows that the skyrmion lattice has a nonzero Hall conductivity. One might then be inclined to take the average value of the fictitious magnetic field and ignore the spatial dependence when calculating the thermal Hall conductivity of the skyrmion lattice. However, we will show shortly that the spatial variation of the fictitious magnetic field is substantial, so that we should take both contributions into account in our analysis. 

To illustrate this point, let us consider the situation in which $R = \pi/\zeta$. In that case $B_z(\rho) = \zeta^2 \cos \theta(\rho)$. The spatial variation is therefore large enough that the fictitious field switches from a negative minimum at $\rho = 0$ to a positive maximum at $\rho = R$. Such large variations have been shown to have a significant influence on the band structure of magnetic lattices.~\cite{Cha94}

For what follows, it will be convenient to formally split the fictitious magnetic vector potential in two parts, ${\bf A} ({\bf r}) = {\bf A}_0({\bf r}) + {\bf A}'({\bf r})$, where ${\bf A}_0({\bf r})$ describes the contribution from the homogeneous nonzero average fictitious magnetic flux, and ${\bf A}'({\bf r})$ the periodic contribution with zero average (we work in the Landau gauge)
\begin{eqnarray}
 {\bf A}_0 ({\bf r}) & = & -B_0y {\bf x}, \nonumber\\
{\bf A}'({\bf r}) & = & \sum_{\tau,\eta} \left[ A_x(\tau,\eta){\bf x}+A_y(\tau,\eta){\bf y}\right] e^{i \left(\tau {\bf k}_1 + \eta {\bf k}_2\right) \cdot {\bf r}}.
\end{eqnarray}
Here, $B_0 = 8\pi/(\sqrt{3}a^2)$ is the average fictitious magnetic field, and ${\bf k}_1 = (2 \pi /a) ({\bf x} - {\bf y}/\sqrt{3})$ and ${\bf k}_2 = (2 \pi /a) (2/\sqrt{3}){\bf y}$ are the basis vectors of the reciprocal lattice, such that the periodic part of the fictitious vector potential satisfies $ {\bf A}'({\bf r} + {\bf a}_1) = {\bf A}'({\bf r} + {\bf a}_2) = {\bf A}'({\bf r})$. Such spatially varying magnetic fields are known to give rise to a finite Hall conductivity, even in the absence of a nonzero average.~\cite{Hal88}

\section{Thermal Hall conductivity of the skyrmion lattice}
\label{Sec:IV}
Since the magnetic excitations of the skyrmion lattice can be described by a free bosonic Hamiltonian with a spatially varying fictitious magnetic field with on average two magnetic flux quanta per unit cell and the same symmetry as the skyrmion lattice, the eigenstates of the skyrmion lattice are magnetic Bloch states. In Sec. \ref{sec:diag} we will determine the excitation spectrum and explicit form of these states. In Sec. \ref{sec:THall} we will show how the thermal Hall conductivity of the skyrmion lattice is determined by the Berry curvature of these magnetic Bloch states.
\subsection{Diagonalization}
\label{sec:diag}
To find the elementary excitations of the skyrmion lattice, we need to diagonalize the Hamiltonian $\mathcal{H}$ in Eq. (\ref{eq:Hbos}) with the fictitious magnetic vector potential given in Eq. (\ref{eq:Askyr}). We do this by numerically diagonalizing the matrix that results from rewriting $\mathcal{H}$ in the basis of the Landau levels that describe excitations with the appropriate symmetry in the presence of the fictitious magnetic vector potential ${\bf A}_0({\bf r})$ only. Our derivation follows that of Ref. \onlinecite{Cha94}, with the difference that we consider the case with two flux quanta instead of one flux quantum per unit cell. 

The eigenstates of a free system of dimensions $L\times L$ with only a homogeneous magnetic field $B_0{\bf z}$ and without any underlying symmetries are given by
\begin{equation}
 \psi_{nk_x} ({\bf r}) = \frac{N_n}{\sqrt{L}}e^{-ik_xx}\varphi_n(B_0^{\frac{1}{2}}y+B_0^{-\frac{1}{2}}k_x),
\end{equation}
where $N_n = \frac{1}{\sqrt{2^nn!}}\left(\frac{B_0}{\pi}\right)^{\frac{1}{4}}$ and $\varphi_n(x) = e^{-x^2/2}H_n(x)$, with $H_n(x)$ the $n$-th Hermite polynomial. The corresponding energies are $E_n = 2 J B_0 (n+1/2)$.
To account for the presence of the triangular lattice, and the fact that every unit cell contains two flux quanta, we need to find the most general linear combination of eigenstates that satisfies
\begin{eqnarray}
 \hat{M}_{{\bf a}_1} \psi_{nm{\bf k}}({\bf r}) & = & e^{ik_1a} \psi_{nm{\bf k}}({\bf r}), \nonumber\\
\hat{M}_{{\bf a}_2} \psi_{nm{\bf k}}({\bf r}) & = & e^{ik_2a} \psi_{nm{\bf k}}({\bf r}).
\end{eqnarray}
Here, $k_1$ and $k_2$ are defined such that $(2\pi/a){\bf k} = k_1 {\bf k}_1 + k_2 {\bf k}_2$. Furthermore, ${\bf k}$ is restricted to lie within the first Brillouin zone. We will discuss the origin of the quantum number $m$ later. We have to work with magnetic translation operators $\hat{M}_{{\bf a}_{1,2}}$ since the canonical momentum is no longer a good quantum number in the presence of the vector potential ${\bf A}_0({\bf r})$. These magnetic translation operators are defined as $\hat{M}_{{\bf a}_1} = \hat{T}_{{\bf a}_1}$ and $\hat{M}_{{\bf a}_2} = \exp[-i(4\pi/a)x]\hat{T}_{{\bf a}_2}$, where $\hat{T}_{{\bf a}_{1,2}}$ are the usual translation operators. The appropriate eigenstates are then given by
\begin{eqnarray}
 \psi_{nm{\bf k}}({\bf r}) & = & \sum_{l=-\infty}^\infty (-1)^{(l+\frac{m}{2})(l+\frac{m}{2}-1)}e^{-i(l+\frac{m}{2})(\frac{k_1}{2}-k_2)a}\nonumber\\
& & \times\psi_{n,-k_1-(l+\frac{m}{2})\frac{4\pi}{a}}.
\label{eq:psigen}
\end{eqnarray}
The quantum number $m$, which in our case can take values 0 or 1, accounts for the fact that in the presence of a natural number $p$ of flux quanta per unit cell each magnetic band will split up in $p$ subbands. These subbands are degenerate for a constant magnetic field but will generally split for a spatially varying magnetic field, as we will see later. The set of wave functions defined in Eq. (\ref{eq:psigen}) constitutes a complete orthonormal basis with triangular symmetry. The eigenfunctions are chosen in such a way that perturbations in the fictitious magnetic vector potential that are periodic in the triangular lattice are diagonal in the momenta $k_1$ and $k_2$.
\begin{figure}
\centering
\includegraphics[width=0.99\columnwidth]{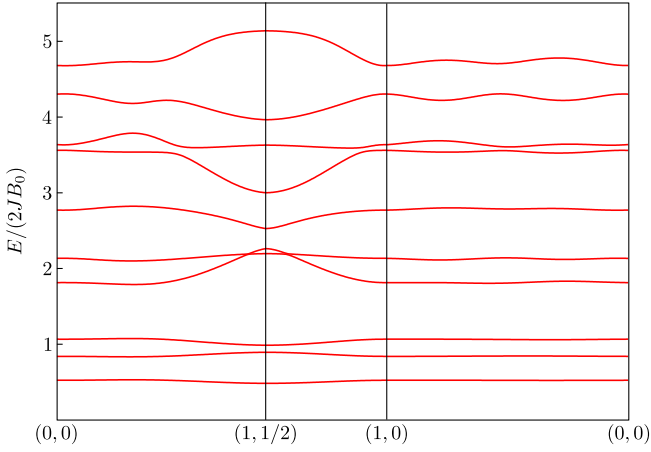}
\caption[]{Band structure of the skyrmion lattice with parameters $R=45$ nm, $\zeta = 70$ $\mu$m$^{-1}$, and $2JB_0/k_B \approx 50$ mK. The labels on the horizontal axis denote $(k_1,k_2)$, with the wave vectors normalized to $2\pi/a$.}
\label{fig:Fig3}
\end{figure}

We are now in a position to calculate the matrix elements of $\mathcal{H}$ with respect to the basis defined by the eigenstates in Eq. (\ref{eq:psigen}). We rewrite $\mathcal{H} = \mathcal{H}_0 + \mathcal{H}_1 + \mathcal{H}_2$, where the subscript denotes the order in which ${\bf A}'({\bf r})$ occurs in the respective term. The matrix elements of $\mathcal{H}_0$ are then trivially given by (we have suppressed the ${\bf k}$ dependence of the eigenstates in our notation)
\begin{equation}
 \la n',m'| \mathcal{H}_0 |n,m \ra = 2 J B_0\left(n + 1/2\right) \delta_{n,n'}\delta_{m,m'}.
\end{equation}
The matrix elements of $\mathcal{H}_1$ are given by
\begin{multline}
\la n',m'| \mathcal{H}_1 |n,m \ra_{n'\ge n} = J \sum_{\tau,\eta} \delta^{(\textrm{mod } 2)}_{m'-m,\tau} B(\tau,\eta)\\
 \times \left[L_n^{n'-n}(z_{\tau \eta}) - \left(\frac{n+n'}{z_{\tau \eta}}L_n^{n'-n}(z_{\tau \eta})-\frac{2n'}{z_{\tau \eta}}L_{n-1}^{n'-n}(z_{\tau \eta})\right)\right]\\
\times (-1)^{m\eta} G_{n' n }(\tau,\eta),
\end{multline}
and the matrix elements of $\mathcal{H}_2$ by
\begin{multline}
\la n',m'| \mathcal{H}_2 |n,m \ra_{n'\ge n} = J \sum_{\tau',\eta',\tau,\eta}\delta^{(\textrm{mod } 2)}_{m'-m,\tau'+\tau} \\
\times \left[A_x(\tau',\eta')A_x(\tau,\eta)+A_y(\tau',\eta')A_y(\tau,\eta)\right]\\
\times (-1)^{m(\eta'+\eta)} G_{n' n }(\tau'+\tau,\eta'+\eta).
 \end{multline}
We defined the function
\begin{eqnarray}
 G_{n' n }(\tau,\eta) & = & \left(\frac{n!}{n'!}\right)^{1/2}(\sqrt{2/B_0}\pi)^{n'-n}\left[i\frac{2\eta-\tau}{\sqrt{3}a} - \frac{\tau}{a}\right]^{n'-n} \nonumber\\
& & \times e^{-z_{\tau \eta}/2}e^{\pi i \tau \eta/2}e^{i\eta k_1a/2}e^{i\tau(k_2 a+\pi)/2}.
\end{eqnarray}
Furthermore, we defined $z_{\tau\eta} = (2\pi/\sqrt{3})(\tau^2-\tau\eta+\eta^2)$. The function $L_n^\alpha(x)$ is the associated Laguerre polynomial. The function $\delta^{(\textrm{mod } 2)}_{i,j}$ is defined as $\delta^{(\textrm{mod } 2)}_{i,j}=1$ when $i=j(\textrm{mod }2)$, and $\delta^{(\textrm{mod } 2)}_{i,j}=0$ otherwise.
The first ten subbands of the band structure of the skyrmion lattice with parameters $2JB_0/k_B \approx 50$ mK, $R = 45$ nm, and $\zeta = 70$ $\mu$m$^{-1}$ (similar values to those found in Ref. \onlinecite{Yu10}) are given in Fig. \ref{fig:Fig3}. In our numerical calculation we used the fact that the coupling between two band decays superexponentially [to be precise, it decays as $\sqrt{(n!/n'!)}$], so that only a limited number of bands have to be taken into account. It is seen that the inclusion of the spatially varying fictitious magnetic field has a pronounced effect, leading both to different splittings of the different subbands, as well as substantial broadening of the subbands. From Fig. \ref{fig:Fig3} it is seen that the typical level splitting between magnetic subbands is 50 mK, which sets the temperature scale on which the system is in the quantum Hall regime. Systems with larger ratio $\Gamma_{\textrm{DM}}^2/J$ will display quantum Hall behavior at higher temperatures. We note that finite Gilbert damping $\alpha$ will broaden the different magnetic subbands by an amount $(\Delta \omega/\omega) = 2 \alpha$. Eventually this will destroy the visibility of individual subbands. However, since the Gilbert damping is around  $\alpha \sim 10^{-3}$ in a range of different materials, this only becomes problematic at high magnetic subbands.

We note that within our model we do not find the expected Goldstone modes associated with the skyrmion lattice.~\cite{Pet11} We argue that this is  due to our adiabatic assumption, which breaks down for the smallest wave vectors. Assuming a quadratic dispersion for the magnons, we can estimate the magnitude $|{\bf k}_m|$ of the characteristic wave vector of the magnons that make up the lowest magnetic subband as $J|{\bf k}_m|^2 = JB_0$, which leads to a typical magnon wave length $\lambda_m \sim a$. The wave vector $|{\bf k}_m|$ increases for higher subbands. Since the accuracy of our model increases with increasing wave vector, our description improves for higher magnetic subbands.

In the next section we will investigate the effect of the finite bandwidth of the magnetic subbands on the thermal Hall conductivity of the skyrmion lattice.

\subsection{Thermal Hall conductivity}
\label{sec:THall}
\begin{figure}
\centering
\includegraphics[width=0.99\columnwidth]{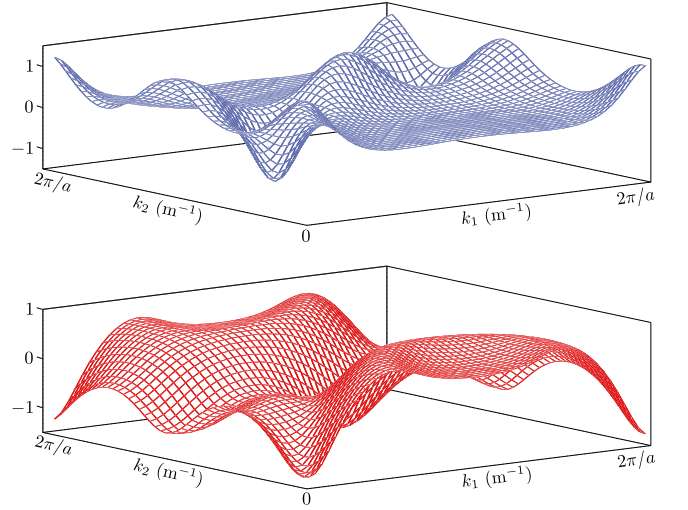}
\caption[]{Berry curvature of the two highest magnetic subbands in Fig. \ref{fig:Fig3} in a single Brillouin zone. The subband corresponding to the top figure does not carry a net curvature, the bottom figure carries 2$\pi$.}
\label{fig:Fig4}
\end{figure}
It is well known~\cite{Cha96} that the semiclassical dynamics of a wave packet in the basis of the magnetic Bloch states $u_{n{\bf k}}({\bf r}) = e^{-i{\bf k}\cdot {\bf r}}\psi_{n{\bf k}}({\bf r})$ is described by
\begin{eqnarray}
 \dot{{\bf r}} = \partial_{{\bf k}} E_n({\bf k}) - \dot{{\bf k}} \times {\bf \Omega}_n({\bf k}) \textrm{ and } \hbar \dot{{\bf k}} = 0.
\label{eq:EOMWP}
\end{eqnarray}
We have assumed here that there are no electric fields present and that the states $u_{n{\bf k}}({\bf r})$ are the eigenstates of the Hamiltonian $\mathcal{H}$, including the fictitious magnetic vector potential ${\bf A}({\bf r})$. ${\bf \Omega}_n({\bf k})$ is the Berry curvature of the $n$th magnetic Bloch band. Since we consider a two-dimensional system, only its $z$ component is relevant. It is given by
\begin{equation}
\Omega_n({\bf k}) = 2\textrm{Im} \left[\left\la \frac{u_{n{\bf k}}({\bf r})}{\partial k_x}\left|\frac{u_{n{\bf k}}({\bf r})}{\partial k_y}\right.\right\ra\right].
\end{equation}
For the skyrmion lattice, the magnetic Bloch states are given by
\begin{equation}
 u_{n {\bf k}}({\bf r}) = e^{-i{\bf k}\cdot {\bf r}} \sum_{n',m'} c^n_{n'm'{\bf k}} \psi_{n'm'{\bf k}}({\bf r}).
\end{equation}
The weights $c^n_{n'm'{\bf k}}$ follow from the diagonalization performed in Sec. \ref{sec:diag}. It should be noted that every completely filled subband carries a total Berry curvature that is a multiple of $2\pi$, in accordance with the quantization of the Thouless-Kohmoto-Nightingale-den Nijs (TKNN) invariant.~\cite{Tho82}

Matsumotu and Murakami have shown~\cite{Mat11} that the thermal Hall conductivity for a system described by Eq. (\ref{eq:EOMWP}) is given by
\begin{equation}
 \kappa_{xy} = \frac{k_B^2T}{\hbar V} \sum_{n, {\bf k}} c_2(\rho_{n{\bf k}})\Omega_n({\bf k}).
\label{eq:kappa}
\end{equation}
Here $c_2(\rho) = (1+\rho)\left(\log\frac{1+\rho}{\rho}\right)^2-(\log\rho)^2-2\textrm{Li}_2(-\rho)$ describes the effect of the thermal distribution of the magnons, and $\rho(\eps) = (\exp^{\beta(\eps-\mu)}-1)^{-1}$ is the Bose-Einstein distribution function.

Figure \ref{fig:Fig4} shows the Berry curvature for the two highest magnetic subbands in Fig. \ref{fig:Fig3}. Together with the band structure given in Fig. \ref{fig:Fig3}, the Berry curvature completely determines the thermal Hall conductivity,  as can be seen from Eq. (\ref{eq:kappa}). The position and width of the magnetic subbands determine at which energies states become available for thermal transport; the Berry curvature determines the extent to which these states contribute to the thermal Hall conductivity.

We have shown then that the spatially varying fictitious magnetic field gives rise to a nontrivial structure of the Berry curvature as well as a broadening of the magnetic subbands. As follows from Eq. (\ref{eq:kappa}), the combination of these two effects may be studied experimentally by measuring $\kappa_{xy}$ as a function of temperature. As the temperature increases, the thermal distribution of the bosonic magnons broadens, which enables the higher bands to contribute to thermal transport.

Alternatively, one can probe the chiral subgap edge states (above the bands whose total TKNN number is odd) directly by microwave means. The first magnetic subband corresponds to several gigahertz, which falls within the scope of standard microwave techniques. Excitation of some of the low-energy modes of the skyrmion lattice by microwave radiation has been performed~\cite{Ono12} and analyzed.~\cite{Moch12} The magnonic reflectionless wave guide modes are analogous to those found in photonic crystals~\cite{Hal08} and can provide intriguing spintronics applications. Magnonic edge states have recently been proposed to exist in YIG without any equilibrium spin texture but with an array with an array of Fe-filled pillars, in which ellipticity of magnons due to dipolar interactions can give rise to magnonic bands with a nonzero TKNN invariant.~\cite{Shi12}

\section{Conclusions}
We have studied how fictitious magnetic fields, which are caused by a textured equilibrium magnetization, lead to thermal Hall effects in two-dimensional insulating magnets in which the nontrivial equilibrium magnetization is caused by spin-orbit interaction. We have given a general expression for the fictitious magnetic vector potential and found that it consists of two contributions: a geometric texture contribution and a contribution due to the original spin-orbit term in the free energy. We have shown that both contributions are generally of the same order of magnitude.

We have derived the relevant properties of the two ground states of interest to us (the spiral state and the skyrmion lattice state) in the phase diagram of a two-dimensional non-itinerant ferromagnet with nonzero Dzyaloshinskii-Moriya interaction. We have found that a system which has the spiral state as magnetic ground state does not have a finite thermal Hall conductivity. However, we predicted that certain domain wall structures do display thermal Hall effects.

We have numerically diagonalized the Hamiltonian describing the triangular skyrmion lattice. We found that due to the spatially varying fictitious magnetic vector potential, the excitation spectrum consists of broadened magnetic subbands. We have calculated the Berry curvature of the magnetic subbands and showed that the Berry curvature in combination with the excitation spectrum completely determines the thermal Hall conductivity of the skyrmion lattice. At present, we are only able to capture the contribution to the thermal Hall conductivity from higher magnetic subbands, as well as thermal- or microwave transport through the associated edge states. In order to properly describe the lowest subbands, our model has to be amended to capture non-adiabatic magnon-transport effects, in the way described at the end of Sec. \ref{Sec:II}.
\section{Acknowledgements}
This work has been supported by the Swiss NSF, the NCCR Nanoscience Basel (K.v.H. and D.L.), the NSF under Grant No. DMR-0840965, and DARPA (Y.T.). K.v.H. is grateful for hospitality at the University of California, Los Angeles, where part of this work has been carried out.
\appendix


\begin{thebibliography}{99}

\bibitem{Trauz08}
B. Trauzettel, P. Simon, and D. Loss, Phys. Rev. Lett. {\bf 101}, 017202 (2008).

\bibitem{Mei03}
F. Meier and D. Loss, Phys. Rev. Lett. {\bf 90}, 167204 (2003).

\bibitem{Uch10}
K. Uchida, J. Xiao, H. Adachi, J. Ohe, S. Takahashi, J. Ieda, T. Ota, Y. Kajiwara, H. Umezawa, H. Kawai, G. E. W. Bauer, S. Maekawa, and E. Saitoh, Nature Materials {\bf 9}, 894 (2010).

\bibitem{Kaj10}
Y. Kajiwara, K. Harii, S. Takahashi, J. Ohe, K. Uchida, M. Mizuguchi, H. Umezawa, H. Kawai, K. Ando, K. Takanashi, S. Maekawa, and E. Saitoh, Nature {\bf 464}, 262 (2010).

\bibitem{Wei12}
M. Weiler, M. Althammer, F. D. Czeschka, H. Huebl, M. S. Wagner, M. Opel, I.-M. Imort, G. Reiss, A. Thomas, R. Gross, and S. T. B. Goennenwein, Phys. Rev. Lett. {\bf 108}, 106602 (2012).

\bibitem{San11}
C. W. Sandweg, Y. Kajiwara, A. V. Chumak, A. A. Serga, V. I. Vasyuchka, M. B. Jungfleisch, E. Saitoh, and B. Hillebrands, Phys. Rev. Lett. {\bf 106}, 216601 (2011). 

\bibitem{Hoo11}
K. A. van Hoogdalem and D. Loss, Phys. Rev. B {\bf 84}, 024402 (2011) and {\it ibid.} {\bf 85}, 054413 (2012) 

\bibitem{Kat10}
H. Katsura, N. Nagaosa, and P. A. Lee, Phys. Rev. Lett. {\bf 104}, 066403 (2010).

\bibitem{Ono10}
Y. Onose, T. Ideue, H. Katsura, Y. Shiomi, N. Nagaosa, and Y. Tokura, Science {\bf 329}, 297 (2010).

\bibitem{She04}
D. D. Sheka, I. A. Yastremsky, B. A. Ivanov, G. M. Wysin, and F. G. Mertens, Phys. Rev. B {\bf 69}, 054429 (2004).

\bibitem{Dug05}
V. K. Dugaev, P. Bruno, B. Canals, and C. Lacroix, Phys. Rev. B {\bf 72}, 024456 (2005).

\bibitem{Li11}
Y.-Q. Li, Y.-H. Liu, and Y. Zhou, Phys. Rev. B {\bf 84}, 205123 (2011).

\bibitem{Han10}
J. H. Han, J. Zang, Z. Yang, J.-H. Park, and N. Nagaosa, Phys. Rev. B {\bf 82}, 094429 (2010). 

\bibitem{Kov09}
A. A. Kovalev and Y. Tserkovnyak, Phys. Rev. B {\bf 80}, 100408(R) (2009).

\bibitem{Hat}
M. Hatami, G. E. W. Bauer, Q. Zhang, and P. J. Kelly, Phys. Rev. Lett. {\bf 99} 066603 (2007) and {\it ibid.}, Phys. Rev. B {\bf 79}, 174426 (2009).

\bibitem{Ber}
L. Berger, J. Appl. Phys. {\bf 58}, 450 (1985), S. U. Jen and L. Berger, J. Appl. Phys. {\bf 59}, 1278 (1986),  and {\it ibid}, 1285 (1986).

\bibitem{Bych84}
Yu. A. Bychkov and E. I. Rashba, J. Phys. C: Solid State Phys. {\bf 17}, 6039 (1984).

\bibitem{Tse11}
Y. Tserkovnyak and D. Loss, Phys. Rev. Lett. {\bf 108}, 187201 (2012).

\bibitem{DM58}
See I. E. Dzyaloshinskii, J. Phys. Chem. Solids {\bf 4}, 241 (1958) and T. Moriya, Phys. Rev. {\bf 120}, 91 (1960).

\bibitem{Kov12}
A. A. Kovalev and Y. Tserkovnyak, Europhys. Lett. {\bf 97}, 67002 (2012).

\bibitem{Bog94}
A. Bogdanov and A. Hubert, J. of Mag. and Mag. Mat. {\bf 138}. 255 (1994).

\bibitem{Yu10}
X. Z. Yu, Y. Onose, N. Kanazawe, J. H. Park, J. H. Han, Y. Matsui, N. Nagaosa, and Y. Tokura, Nature {\bf 465}, 901 (2010).

\bibitem{Nog10}
A. Nogaret, J. Phys.: Condens. Matter {\bf 22}, 253201 (2010).

\bibitem{Ibr95}
I. S. Ibrahim and F. M. Peeters, Phys. Rev. B {\bf 52}, 17321 (1995).

\bibitem{Mat00}
A. Matulis and F. M. Peeters, Phys. Rev. B {\bf 62}, 91 (2000).

\bibitem{Rad11}
L. Radzihovsky and T. C. Lubensky, Phys. Rev. E {\bf 83}, 051701 (2011).

\bibitem{Cha94}
M. C. Chang and Q. Niu, Phys. Rev. B {\bf 50}, 10843 (1994).

\bibitem{Hal88}
F. D. M. Haldane, Phys. Rev. Lett. {\bf 61}, 2015 (1988).

\bibitem{Pet11}
O. Petrova and O. Tchernyshyov, Phys. Rev. B {\bf 84}, 214433 (2011).

\bibitem{Cha96}
M.-C. Chang and Q. Niu, Phys. Rev. B {\bf 53}, 7010 (1996).

\bibitem{Tho82}
D. J. Thouless, M. Kohmoto, M. P. Nightingale, and M. den Nijs, Phys. Rev. Lett. {\bf 49}, 405 (1982).

\bibitem{Mat11}
R. Matsumoto and S. Murakami, Phys. Rev. Lett. {\bf 106}, 197202 (2011).

\bibitem{Ono12}
Y. Onose, Y. Okamura, S. Seki, S. Ishiwata, and Y. Tokura, Phys. Rev. Lett. {\bf 109}, 037603 (2012).

\bibitem{Moch12}
M. Mochizuki, Phys. Rev. Lett. {\bf 108}, 017601 (2012).

\bibitem{Hal08}
F. D. M. Haldane and S. Raghu, Phys. Rev. Lett. {\bf 100}, 013904 (2008).

\bibitem{Shi12}
R. Shindou, R. Matsumoto, and S. Murakami, arXiv:1204.3349v1 [cond-mat.mes-hall].
\end{thebibliography}
\end{document}